# Scalable CVD Graphene Field-Effect Transistor Platform for Viral Detection: Application to COVID-19

Leonardo Martini[a], Ylea Vlamidis[a,b], Ileana Armando[a,c], Domenica Convertino[a], Vaidotas Mišeikis[a,b], Valerio Voliani[a,d], and Camilla Coletti[a,b]

a. Center for Nanotechnology Innovation@NEST, Istituto Italiano di Tecnologia, Piazza San Silvestro, 12 56126 Pisa, Italy
b. Graphene Labs, Istituto Italiano di Tecnologia, Via Morego 30, 16163 Genova, Italy
c. Center for Material Interface@SSSA, Istituto Italiano di Tecnologia, 56025 Pontedera, Italy
d. Department of Pharmacy, School of Medical and Pharmaceutical Sciences, University of Genoa, Viale Cembrano 4, Genoa 16148, Italy

*Abstract*— The rapid and global spread of coronavirus disease 2019 (COVID-19), caused by the severe acute respiratory syndrome coronavirus 2 (SARS-CoV-2), underscored the urgent need for fast, reliable, and adaptable diagnostic tools capable of responding to current and future viral threats. Early diagnosis is key to limiting transmission, and biosensors based on nanomaterials offer promising solutions for accurate and rapid bioanalyte detection.

In this work, we present a scalable matrix of graphene-based field-effect transistors (GFETs) for the direct and rapid detection of the SARS-CoV-2 spike protein. High-quality graphene is functionalized in a single step with ACE2–His, enabling detection of the spike protein with a limit of detection as low as 1 fg/mL in phosphate-buffered saline (PBS). A robust statistical analysis, based on measurements from approximately 70 devices per analyte concentration, demonstrates the reproducibility and reliability of the platform. This label-free, scalable, and reproducible COVID-19 antigen sensor can be readily adapted to detect emerging SARS-CoV-2 variants or other viral pathogens, offering a flexible approach for future diagnostic applications.

## I. INTRODUCTION

In late 2019, a coronavirus known as SARS-CoV-2 emerged in Wuhan, China [1] leading to the global COVID-19 pandemic, causing over millions of cases and deaths worldwide [2]. The pandemic has underscored the critical need for rapid, sensitive, and reliable diagnostic tools. While traditional methods like RT-PCR are highly accurate, they require specialized facilities and trained personnel, limiting their widespread use. Antigen tests offer faster results but with reduced sensitivity. Similarly, antibody tests, which detect past infections, cannot diagnose active cases as antibodies take time to develop [3]. These limitations have prompted new research into the development of innovative and portable diagnostic platforms.

Coronaviruses are enveloped viruses with positive sense single–strained RNA genome that can infect birds and mammalians. The clinical manifestation of the disease usually starts after less than a week, consisting of upper respiratory tract infections with a spectrum of clinical manifestations that typically include fever, cough and fatigue, sometimes with pulmonary involvement. The genome of SARS-CoV-2 has been fully sequenced and shows similarity with other coronaviruses causing respiratory diseases such as SARS-CoV-2 [4]. The coronavirus virion particle is typically round or multi–shaped. On the virion surface are located several proteins, including the spike glycoprotein (S), which plays a key role in SARS-CoV infection. S glycoprotein consists of two subunits: S1 containing a receptor binding domain (RBD), which enables the attachment of the virus to the host cells; S2 that is involved in the subsequent fusion of the virus and host membrane [5], [6]. In addition to the specific S protein, coronavirus genomes generally encode three additional structural proteins: the membrane protein (M), the envelope protein (E) and the Nucleocapsid (N) that is involved in the viral replication [7]. The virus infects lung alveolar epithelial cells using receptor–mediated endocytosis via the angiotensin-converting enzyme II (ACE2) as an entry receptor [8], [9]. In particular, the first step of infection involves the recognition of the ACE2 receptor by the membrane spike glycoprotein of SARS-CoV-2 binding of C-terminal domain (i.e. RBD of the envelope–embedded S1 protein) of the SARS-CoV-2 to ACE2 [10]. Since the initial outbreak, multiple SARS-CoV-2 variants with mutations in the spike protein have emerged, affecting transmissibility and immune recognition. This underscores the importance of flexible sensing platforms that can be rapidly functionalized with variant-specific antibodies to track the spread of new, potentially more aggressive strains.

Biosensors have gained significant attention for their ability to detect SARS-CoV-2 and other pathogens [8]. Various biosensor technologies, including electrochemical, optical, and electrical sensors, provide rapid detection and enable mass screening, particularly in high-risk settings [11], with recent studies highlighting nanotechnology–based using gold nanoparticles, carbon nanotubes [12], quantum dots, and also graphene [13]. Among these, graphene-based field-effect transistors (GFETs) stand out due to their exceptional electrical properties, high surface-to-volume ratio, and biocompatibility, making them ideal for label-free, real-time monitoring of biomolecules at low concentrations [14][15]. These sensors offer remarkable sensitivity, enabling the detection of low concentrations of biomolecules such as proteins[16], biomolecules[17], glucose[18], DNA[19], and viruses[15], [20]. Furthermore, their versatility, combined with the ability to integrate with other technologies, positions GFETs as powerful tools for a wide range of biomedical applications, from disease detection to environmental monitoring [19], [21], [22]. Graphene-based sensors have been proven as working devices in the detection of several biomarkers, from beta-2-microglobiline in biological samples [16], to the detection of simpler molecules in a graphene CMOS-compatible device[23]. Recent advancements in GFET-based biosensors for SARS-CoV-2 detection demonstrate their potential for high sensitivity and rapid, cost-effective diagnostics, with the possibility of future portable, wearable applications. For instance, several devices for rapid detection of COVID-19 based on GFET has proven detectivities in the fg/ml range [20], [24], [25], [26]. However, up to now, there is a lack of studies focusing on higher statistics with a significant number of devices. Clearly, such investigations are needed to effectively assess the viability of this technology and define a pathway towards its realistic implementation.

In this work, we report the development of a graphene-based biosensor for the rapid detection of the COVID-19 RBD receptor with a significant number of devices realized on high-quality single-crystal graphene grown in matrices [27]. The graphene is non-covalently functionalized with cellular receptor ACE2 for the recognition and binding of the RBD of SARS-CoV-2 spike protein, as proposed by *Seo et al.* [20]. The biosensor is designed with an array of 100 micro transistors, including functionalized and control groups. This extensive sampling enables robust statistical analysis, improves confidence in signal trends, and allows for a more accurate assessment of variability, reproducibility, and saturation effects. The devices are fabricated on graphene single–crystal matrices grown by chemical vapour deposition (CVD). The adoption of single-crystals ensures high sensitivity of the devices and the synthetic approach guarantees scalability [27]. The biosensor functions as a solution-gated transistor, where the electrical double layer formed by the electrolyte ions serves as the gate. The detectivity of the demonstrated device is as low as 1fg/mL. Owing to the ease of surface functionalization, the GFET platform can be readily adapted for the detection of emerging SARS-CoV-2 variants and other viral pathogens, making it a versatile tool for future diagnostic challenges.

## II. MATERIAL AND METHODS

### A. *Graphene growth, transfer and fabrication*

Matrices of single–crystal graphene were grown in a deterministic manner via chemical vapor deposition (CVD) on electropolished copper (Alfa Aesar 99.8%) in a commercially–available cold–wall CVD reactor (Aixtron 4" BM Pro), as previously reported [28]. The initial step of seeded growth involved electropolishing and patterning of the growth substrate. Nucleation sites were patterned via optical lithography, followed by thermal evaporation of a 25 nm chromium layer and lift-off. Graphene growth was carried out in a commercially available cold-wall reactor (Aixtron 4″ BM Pro) using a low-pressure (25 mbar) CVD process. The substrates were first annealed for 10 minutes in a non-reducing argon atmosphere to preserve the native oxide layer. Growth was then performed at 1060 °C with a gas mixture consisting of 900 standard cubic centimeters per minute (sccm) of argon, 100 sccm of hydrogen ($H_2$), and 1 sccm of methane ($CH_4$). Graphene crystals were then transferred on the target substrate with a deterministic semi–dry procedure. First, graphene on copper was covered with a polymeric membrane of PMMA and a few–millimeters PDMS frame, to ensure mechanical rigidity; the graphene/membrane was then delaminated from the copper foil and transferred on the target substrate [27] commercially–available p-doped silicon (Si) covered with 285 nm silicon dioxide ($SiO_2$)).

Electron beam lithography (EBL) (Zeiss Ultra Plus) was employed to define the active area of the 100 FET devices, which were all fabricated at the same time. To minimize the effect of the contact resistance on the graphene-channel characterization and the Joule heating during the measurement, the devices were designed in a 4-probe configuration [29]. Concurrently, a single Hall bar was designed in each sample, in order to characterize the carrier mobility of the pristine graphene. PMMA 4% in anisole (Allresist) was spin-coated at 4000 rpm for 60 seconds and dried for 5 minutes at 120°C; the polymer was then exposed to a dose of 280 $\mu C/cm^2$ in a Zeiss Gemini scanning electron microscope (SEM) with electron accelerated at 20 kV. Graphene was removed from the exposed areas via reactive ion etching (RIE) (Sistec) for 45 s of $Ar/O_2$ at 5/80 SCCM. Optical lithography was performed in a laser writer MicroWriter ML3 Pro, with 385 nm light source and spatial resolution of 0.7 μm. Microposit S1813 was used as (positive) photoresist, spun at 6000 rpm for 1 minute, and baked for 2 minutes at 110 °C. Thermal metal evaporation (50 nm of gold on top of 5 nm of chromium) was used to pattern the alignment markers and the metallic pads required in the electrical characterization and for source and drain electrodes. Finally, the sample was cleaned in acetone, isopropanol, PMMA–remover (AR 600-71), and deionized water [30]. The size of the active area of the obtained FET devices was 100 x 120 $\mu m^2$ (L x W).

### B. *Graphene functionalization*

The matrix of GFETs was functionalized with the Human ACE2 antibody using the following procedure. The human ACE2 antibody linked with 6X histidine tag at the C–terminus (ACE2–His Tag, amino acids 18–740, purity 90 %, ThermoFisher) was diluted to 100 μg/ml in PBS, pH 7.4. The graphene FET was functionalized by drop casting 40 μL of ACE2–His solution onto the sensor surface and incubated for 2 hours. Finally, the sensors were rinsed in milliQ water. Subsequently, 40 μL of RBD SARS-CoV-2 spike protein (full length SARS-CoV-2 Nucleoprotein/Spike Protein, purity >95\%, ThermoFisher) diluted in PBS, pH 7.4 at different concentrations (10 fg/mL to 0.5 fg/mL) were deposited on the GFET/ACE2 and incubated at room temperature with the antibody for 2 hours, then rinsed in MilliQ water, to observe the antibody–analyte interaction.

### C. *Device characterization*

The devices were characterized via atomic force microscopy (AFM), Raman spectroscopy, and electrical measurements after each step of fabrication and functionalization. Surface roughness analysis of the GFETs was conducted by AFM; the characterization was performed with a Bruker Dimension Icon system employing the ScanAsyst mode. Subsequently, AFM micrographs were analyzed using the software Gwyddion 2.54. Raman spectra were acquired with a Renishaw InVia system equipped with a 532 nm green laser and 1800 l/mm grating. All Raman spectra were collected with a 100x objective lens, at an exposure time of 2 s, 1 accumulation, and laser power of ∼ 1 mW. Mapping images were obtained from 196 spectra (20 × 20 μm$^2$ area on graphene). Intervals between adjacent acquisition points were 1.5 μm both in the horizontal and vertical directions.

### D. Direct current (DC) characterization

Electrical characterization was performed at room temperature and under ambient conditions. The samples were characterized in a probe–station equipped with five manually controlled micropositioners (MPI–corporation MP40 MicroPositioner) each featuring a metallic tip (MPI–corporation 7 μm tungsten tips). Electrical characterization of the GFETs was performed with a Keithley 2450 multimeter, enabling both 2–probe and 4–probe characterization, with either current or potential constant. A constant current was flown between the two external contacts, while the potential drop was measured between the two inner contacts, to eliminate the contribution of the contact resistance in the FET measurement and to better isolate the effect of the external dopant in the device behavior. A second high–tension source–meter was used to provide backgate potential and to control eventual current leakage through the dielectric substrate. Preliminary characterization of the graphene was conducted via ambient-condition Hall measurement, with a commercial 32 mT electromagnet, controlled by a Keithley current source-meter; the Hall measurement was made with constant source-drain current of 100 nA. Out of the 100 fabricated GFETs, we tested ∼70 after functionalization, avoiding those at the physical edge of the drop casted on the sample.

## III. RESULTS AND DISCUSSIONS

The biosensor was realized by adopting an array of graphene single–crystals, grown by CVD and semi-dry transferred on commercial p-doped silicon, as described in the methods and in previous work [28]. Such synthetic and transfer approaches yield scalable monolayer graphene whose mobilities are comparable to those obtained with exfoliated flakes [31], [32]. An array of 100 separate GFETs was fabricated via lithography, with each device placed on a distinct single-crystal (Fig. 1a). To maximise the functionalisation area while avoiding bilayer regions typically present near the graphene nucleation site, each GFET was designed to occupy the largest possible area within the boundaries of the single-crystal domain. Fig. 1b shows an optical micrograph of a representative device. The functionalization process of the GFET is schematically reported in Fig. 1c: following completion of all lithographic steps, functionalization with ACE2-His was performed via drop casting. As a last step, RBD SARS-CoV-2 spike protein was delivered to the sensor again via drop casting. This functionalization approach is reversible via cleaning in DI water and PBS [33], [34] leaving the graphene detector potentially reusable. The biosensor functioned as a solution-gated transistor, where the electrical double layer created by the ions in the electrolyte acts as an external doping source for the GFET.

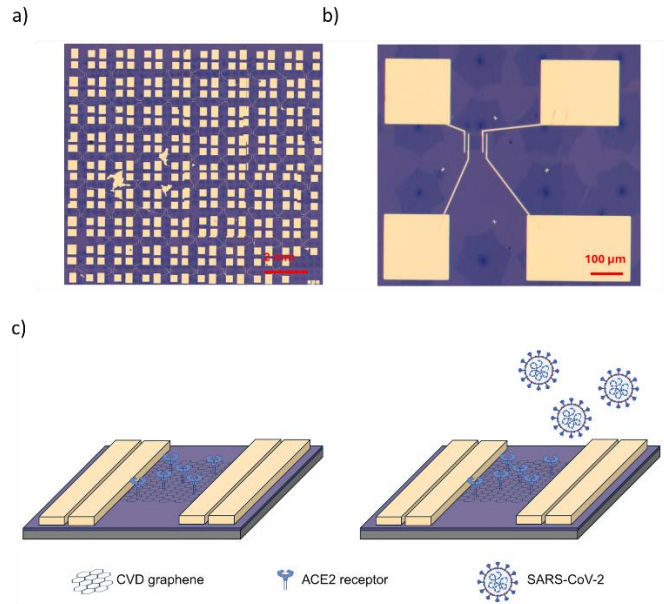

Fig. 1(a) Optical image of an entire array of GFET on a single substrate and (b) of a single device. (c) Schematic representation of the device and of its functionalization: a 4–contact GFET where the antibody ACE2 and the spike protein solutions are drop casted on top. Created with Biorender.com

The morphology of graphene before and after functionalization was studied via AFM. As shown in Fig. 2a, the as-synthesized pristine graphene crystals showed a flat surface, with a root mean square (RMS) roughness ∼0.7 nm. After ACE2-His drop casting, the surface morphology significantly changed and the RMS roughness increased to ∼1.8 nm, as visible in Fig. 2b. We attribute this effect primarily to the inhomogeneity of the ACE coverage after the deposition, since other characterization techniques indicate no particular change in the structural properties of graphene. Following the functionalization with the RBD spike protein (5fg/mL), the roughness was ∼1.4 nm, slightly reduced when compared to the ACE2-functionalized graphene. Similar morphology was observed also for the other tested concentration. This can be attributed to the formation of a homogeneous coating of the device with the spike protein (Fig. 2c), again suggesting that the main cause of the change in the roughness is related to the deposition process.

Raman spectra of graphene were acquired to characterize graphene at each stage of functionalization and are reported in Fig. 3a-c for the case of 5 fg/mL spike protein concentration. Panel d shows representative spectra of the same graphene crystal: in black, graphene after fabrication; in red, after ACE-2 functionalization; and in blue, after spike protein drop-casting. In no case was observed the characteristic D-peak indicative of sp$^3$ bonds [35]. This confirms the high crystallinity of the pristine graphene and further suggests that the drop-casting functionalization steps did not induce covalent modification of the graphene lattice. Both 2D and G peaks can be fitted with a single Lorentzian function, as expected for a single-layer graphene [36]. The average G-peak position is found at 1581.2±1.3 cm$^{-1}$ for the uncovered graphene after the device fabrication, that changes to 1581.4±1.1 cm$^{-1}$ after the ACE-2 functionalization and to 1582.3±1.4 cm$^{-1}$ after the spike deposition [37]. The FWHM of the G-peak changes from 13.5±1.4 cm$^{-1}$ in graphene after device fabrication, to 16.0±2.6 cm$^{-1}$ after the ACE-2 functionalization, and back to 11.2±2.2 cm$^{-1}$ in the graphene with the spike solution drop-casted, suggesting a change in the graphene doping level with the process [38]. The position of the 2D-peak is centered at 2673.3±2.6 cm$^{-1}$ without any relevant change between the functionalization steps, compatible with similarly synthesized graphene devices (see panel f) [39]. The FWHM of the 2D-peak is around 25.0±1.5 cm$^{-1}$ for graphene after fabrication and remains the same after the ACE-2 functionalization; those values are compatible with singe crystal graphene on SiO$_2$ substrate [40], [41]. The fact that both the FWHM and the position of the 2D peak do not measurably change after ACE deposition further indicates that the concurrent increase in surface roughness observed via AFM is not related to graphene structural modifications and can be attributed to the antibody distribution on graphene. The FWHM of the 2D-peak is reduced to about 22.7±0.7 cm$^{-1}$ after the spike deposition, indicating that spike coating of the ACE-functionalized graphene could be beneficial for reduction of strain inhomogeneities [42].

Hall bar devices were realized to attest the carrier mobility of graphene prior to functionalization. Fig. 4a reports a representative Hall-measurement of pristine graphene: holes (electron) mobility was found to be ~7500 cm$^2$/Vs (~5000 cm$^2$/Vs) at a carrier concentration of 2x10$^{12}$ cm$^{-2}$ (1x10$^{12}$ cm$^{-2}$). The high quality of the starting materials is expected to increase the sensitivity to external substances. The behaviour of the GFET sensor was assessed by measuring ~70 devices after each functionalization step. Fig. 4b reports representative transfer curves measured for the same device at each stage of functionalization with ACE2 and a spike concentration of 5 fg/ml. It can be clearly seen how the charge neutrality point shifts progressively from positive voltage towards more neutral values after the ACE drop casting, and toward negative values after Spike deposition, indicating a change in the doping level. Indeed, the graphene doping level is affected by the ACE drop casting step, with an electron donor effect as previously reported [26]. This behaviour is attributed to interactions between the imidazole ring of histidine and the graphene surface. At physiological pH (7.4), histidine is predominantly in its neutral form, with the imidazole ring likely acting as an electron donor in its interaction with graphene. A minor fraction of protonated imidazole groups may also be present, potentially contributing to variations in the observed doping level. At the same time the other key parameters of the graphene, as the carrier mobility and the contact resistivity, are not affected by the functionalization. After the spike protein drop casting, the additional shift in charge neutrality point indicates a n-type doping of graphene. Indeed, the Raman analysis of the FWHM of the G peak agrees well with the doping levels measured electrically. In Fig. 4c the charge neutrality point position is shown, measured on more than 70 devices, at each step of functionalization: the presence of the ACE2-His precursor shifts the graphene neutrality point to a lower doping level, after the spike deposition the n-doping effect increases [20].

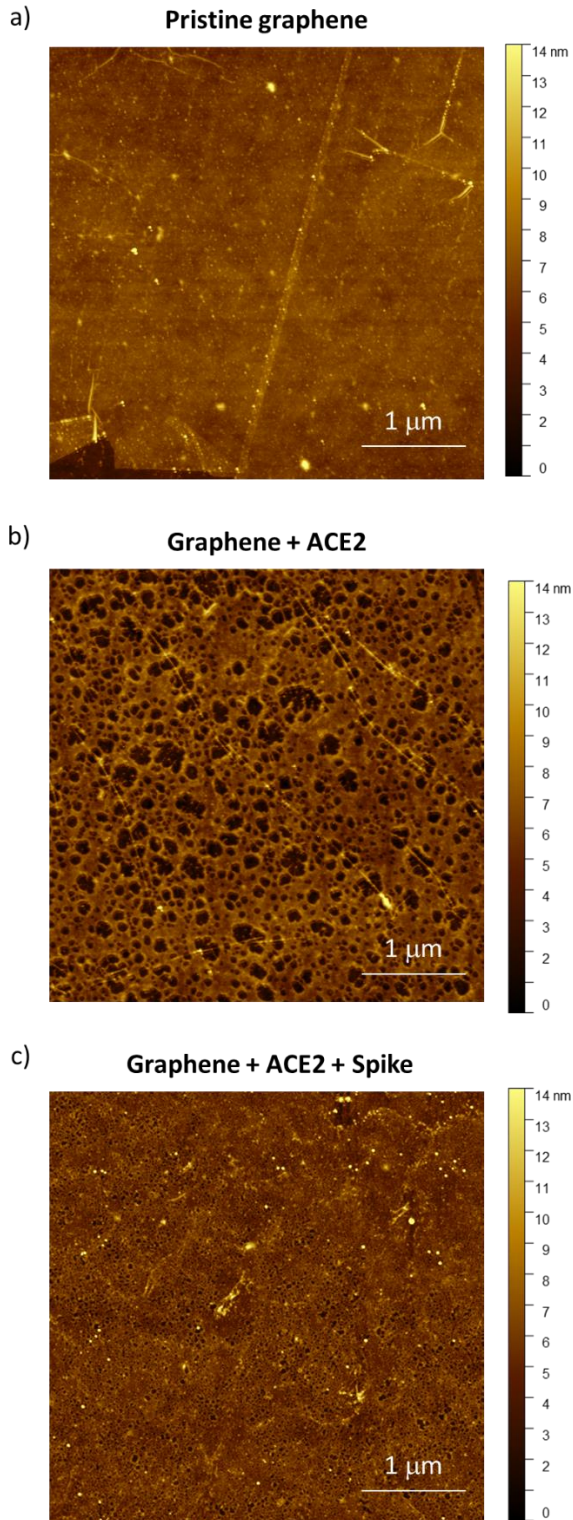

Fig. 2. AFM characterization of pristine (a), ACE2–functionalized (b), and SARS-CoV-2 spike protein–immobilized graphene (c). The pristine graphene presents an average roughness of 0.7 nm, compatible with the substrate, while the roughness increases on the functionalized device.

In this regard, a systematic study of the sensor detection was performed (Fig. 4d): four different concentrations of spike protein, from 0.5 to 10 fg/ml, were deposited on our GFET matrices. In Fig. 4d, the average shift of the charge neutrality point between the deposition of the precursor and the spike-solution at different concentrations are reported. The variance reported is the standard deviation of the measured Dirac point shift for each device. The average value and the variance reported demonstrate a concentration-dependent increase in the doping effect induced by the spike protein ; allowing the possibility to define a minimal concentration sensitivity of 1 fg/mL: above this level, the shift in the charge neutrality can be confidentially attributed to presence of the spike protein, as shown in Fig. 4c, since the change exceeds the natural variability of the doping levels in functionalized graphene devices, that may occur due to the inhomogeneity of the precursor deposition. Finally, our experimental data show a characteristic nonlinear dose-response trend. At very low concentrations (0.5-1 fg/mL), the sensor response exhibits minimal variation in both the mean and standard deviation, suggesting an early onset of surface saturation likely due to the high binding affinity of the receptor layer. However, at higher concentrations (≥10 fg/mL), the response begins to flatten again, consistent with a second saturation regime as the number of available binding sites becomes depleted or the electrical modulation of the graphene channel reaches its upper limit. This double-saturation behavior – both at low and high concentration extremes – supports a sigmoidal binding model, such as the Hill–Langmuir isotherm, reported also in other studies [20], [22], [26].

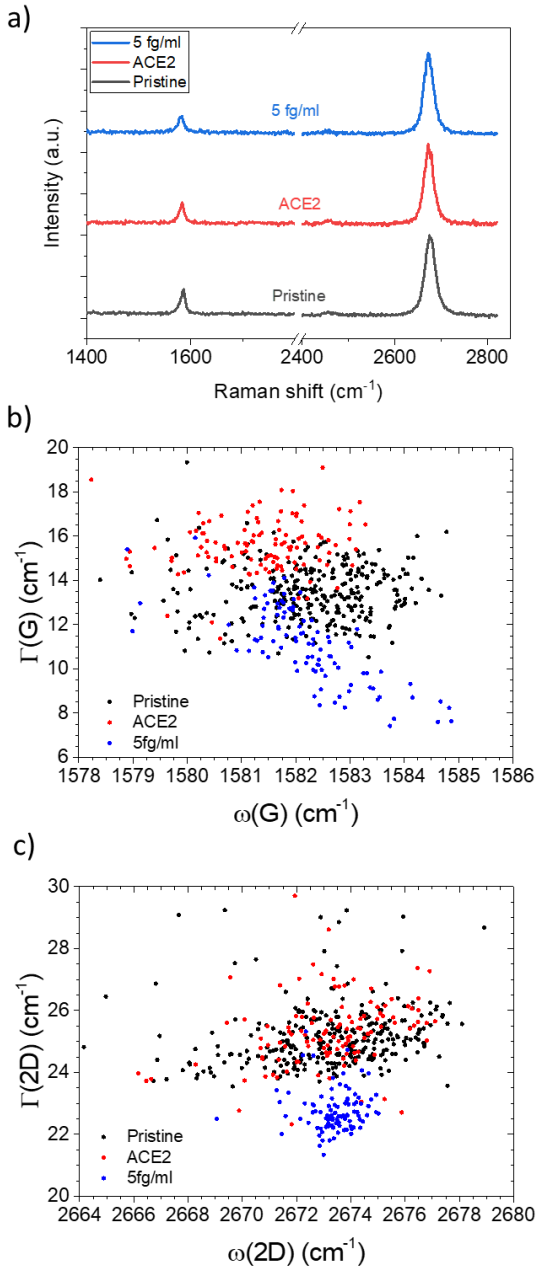

Fig. 3. (a) Comparison of the Raman spectra of the graphene–based device at each functionalization step. In the correlation plot of the peak position and FWHM for the G-peak (b) and 2D-peak (c).

## IV. CONCLUSION

This study reports the development of a highly sensitive electronic sensor for SARS-CoV-2 spike protein detection utilizing high-quality CVD graphene. The fabrication process leverages large-area techniques like chemical vapor deposition and optical lithography, offering cost-effectiveness and scalability for industrial production. Notably, non-covalent functionalization enables targeted chemical responsiveness of the graphene while preserving its exceptional electronic and optical properties. This sensor achieves the remarkable detection limit of 1 fg/mL for the spike protein. This is possible as the graphene adopted presents high quality, with holes (electron) mobility ∼ 7500 cm$^2$/Vs (5000 cm$^2$/Vs), a feature enhancing sensor sensitivity. The measurable shifts in charge neutrality point due to π-π interactions validate the precise detection of biomolecular interactions. We employ a statistically significant number of devices (~70 per concentration), far exceeding the sampling depth reported in comparable studies [20], [25], where statistical detail is often lacking or limited to a handful of devices, increasing the reliability and sensitivity of the biosensor and enabling consistent detection even at trace concentrations. Furthermore, our scalable sensing platform is designed to integrate seamlessly with cost-effective electronic systems, offering a modular and practical solution. Notably, the sensor also shows potential for reusability after multiple washing cycles under controlled temperature conditions, as demonstrated in previous studies [33], [34], further enhancing its practicality and long-term applicability. This study demonstrates the potential of high-sensitivity CVD graphene as a versatile platform for the realization of GFET-based sensors with high sensitivity.

Potentially, these devices could offer the capability to detect diverse viruses, quantify viral loads, and evaluate infectivity.

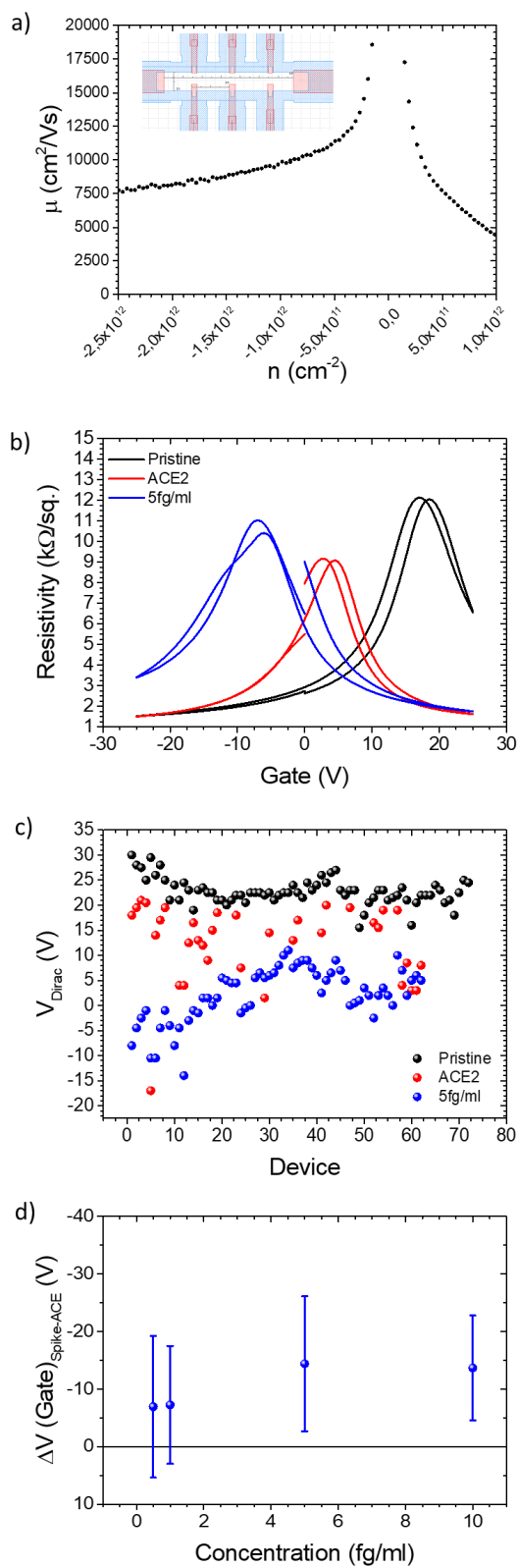

Fig. 4. Electrical characterization of graphene: (a) preliminary Hall measure on the pristine graphene demonstrates the high quality of the material. (b) Transfer curve on a GFET before and after the functionalization with the antibody (ACE2) and the spike (5 fg/ml) solution shows a relevant change in the doping of the material, with a shift of the charge neutrality point position of around 40 V between the pristine graphene and the spike–functionalized one. (c) The distribution of the graphene charge neutrality point position, for an array of 70 devices, before and after the functionalization show a clear shift; the greater variability on the charge neutrality point position after the functionalization may be due to the inhomogeneity of the distribution of the precursor solution. (d) The average shift of the charge neutrality point position, as function of the spike concentration in the solution, shows the sensitivity limit of our system. The statistic has been obtained from the characterization of one sample for each concentration (between 10 and 70 devices each).

ACKNOWLEDGMENT

ACKNOWLEDGMENT

The research leading to these results has received funding from the European Union's Horizon 2020 research and innovation program under grant agreements no. 785219-Graphene Core2 and 881603-Graphene Core3. We acknowledge financial support from NPRR MUR project ECS00000017-THE funded by the European Union – NextGenerationEU. We also thank Professor Gianni Ciofani, coordinator of Center for Materials Interfaces, IIT. Figure 1 was created with BioRender.com.